\documentclass[11pt]{article}
\usepackage{blois,epsfig}
\usepackage{amsmath}
\usepackage{amssymb}

\def\be{\begin{equation}}
\def\ee{\end{equation}}
\def\bea{\begin{eqnarray}}
\def\eea{\end{eqnarray}}

\begin{document}
\vspace*{4cm}
\title{Measuring the Higgs boson couplings}

\author{Michael Rauch}

\address{Institut f\"ur Theoretische Physik, Universit\"at Karlsruhe, \\
  Karlsruhe Institute of Technology, 76128 Karlsruhe, Germany}

\maketitle\abstracts{
Once a Higgs boson has been discovered, it is also important to know
what are its properties, in particular its couplings to the other
particles. At the LHC, there will be many observable channels that can
be used to measure the relevant parameters. Using the SFitter framework,
we project these measurements onto a weak-scale effective theory with
general Higgs boson couplings. Thereby, our analysis profits from the
error treatment and the parameter determination tools already used for
new-physics searches. We use this to study both the individual couplings
and their associated errors as well as correlations between them.
}

\section{Introduction}

Understanding the mechanism of electro-weak symmetry breaking is one of
the main goals of the LHC. In the Standard Model (SM), this is achieved
by the Higgs field, an $SU(2)$ doublet, which obtains a vacuum
expectation value (vev)~\cite{higgs,reviews}. Three degrees of freedom
of this Higgs field are absorbed by the $W$ and $Z$ gauge bosons and
form their longitudinal modes, while the remaining one becomes a scalar,
the Higgs boson. The kinetic terms of the Higgs field in the
Lagrangian lead to $WWHH$ and $ZZHH$ interactions. 
Interactions with fermions are added via Yukawa-type couplings. Once the
Higgs is replaced by its vev in these expressions, we obtain the mass
terms for gauge bosons and fermions. Consequently, the couplings of the
Higgs to the other particles are not free parameters, but proportional
to the known masses and the vev.

The only remaining unknown parameter in the SM is the mass of the Higgs
boson. A lower bound of 114.4 GeV is given by the direct searches from
LEP~\cite{Collaboration:2008ub}, with the latest Tevatron
results approaching a similar sensitivity~\cite{CDF:2011cb} in this
region. Results from the
LHC~\cite{ATLAS-CONF-2011-135,CMS-PAS-HIG-11-022} indicate an upper
limit of 145 GeV. Higgs boson masses above 200 GeV, where allowed
regions from direct searches are still left, are strongly disfavoured by
electro-weak precision data~\cite{Flacher:2008zq}. 
As the Higgs couplings in the SM are completely determined by the known
masses, we can use this theoretical prediction and compare
it~\cite{duehrssen,Lafaye:2009vr,Bock:2010nz,Bonnet:2011yx} with
future LHC measurements of Higgs boson
channels~\cite{atlas_tdr}. For this we assume that the discrete
quantum numbers, like CP structure or
spin~\cite{wbf_vertex}, are identical to the SM value. There are many
models where deviations of the couplings can occur. This includes simple
extensions like a second Higgs doublet, as for example in
supersymmetry~\cite{susy} or also in Higgs portal
models~\cite{Patt:2006fw}, or composite models~\cite{Espinosa:2010vn},
where in a strongly-interacting sector the Higgs emerges as a
pseudo-Goldstone boson.

For this task it is crucial to correctly account for the different types
of errors. As events are counted these statistical errors are
of Poisson type. Systematic errors are correlated between different
measurements and we need to include the full correlation matrix. The
best description for theory errors is the RFit scheme~\cite{ckmfitter},
which takes them as box-shaped. The SFitter tool~\cite{sfitter} allows
us to construct a fully-dimensional log-likelihood map of the parameter
space. To reduce this into plottable one- or two-dimensional
distributions, both Bayesian (marginalisation) and Frequentist
(profile likelihood) techniques are available.

\section{Calculational setup}

As the underlying model we assume the SM with a generalised Higgs sector,
where the Higgs couplings can take arbitrary values. They are
parametrised in the following way: For all particles $j$ with tree-level
couplings to the Higgs, the coupling is modified to
\begin{equation}
g_{jjH} \rightarrow g_{jjH}^{\text{SM}} ( 1 + \Delta_{jjH} ) \ .
\end{equation}
Additionally, there are two important loop-induced couplings, namely
those to photons and gluons. When altering the tree-level couplings,
these will change as well. Additionally, we can introduce dimension-five
contributions, which originate from new particles running in the loop,
e.g.\ the supersymmetric partners in SUSY models. Hence, these couplings
are modified according to
\begin{equation} 
g_{jjH} \rightarrow g_{jjH}^{\text{SM}} ( 1 + \Delta_{jjH}^{\text{SM}} +
\Delta_{jjH}) \ ,
\end{equation} 
where $\Delta_{jjH}^{\text{SM}}$ denotes the contribution from modified
tree-level couplings and $\Delta_{jjH}$ the additional dimension-five
part. We also take the masses of the top and bottom
quark and the Higgs boson as free parameters, as they have, or will
have, significant experimental errors. 
The total width of the Higgs boson is taken as the sum over all SM
particle partial widths.

The LHC measurement channels which enter our analysis are derived from
an ATLAS simulation and assume an integrated luminosity of 30 fb$^{-1}$
at a centre-of-mass energy of 14 TeV~\cite{Lafaye:2009vr,duehrssennote}.
We then smear the simulated experimental results in 10000 toy
experiments and fit the resulting Higgs couplings. From the distribution
we can then read off the errors on the determination.  
More details on the setup and the procedure can be found in
Ref.~\cite{Lafaye:2009vr}.

\section{Results}

\begin{figure}
\centering
\psfig{figure=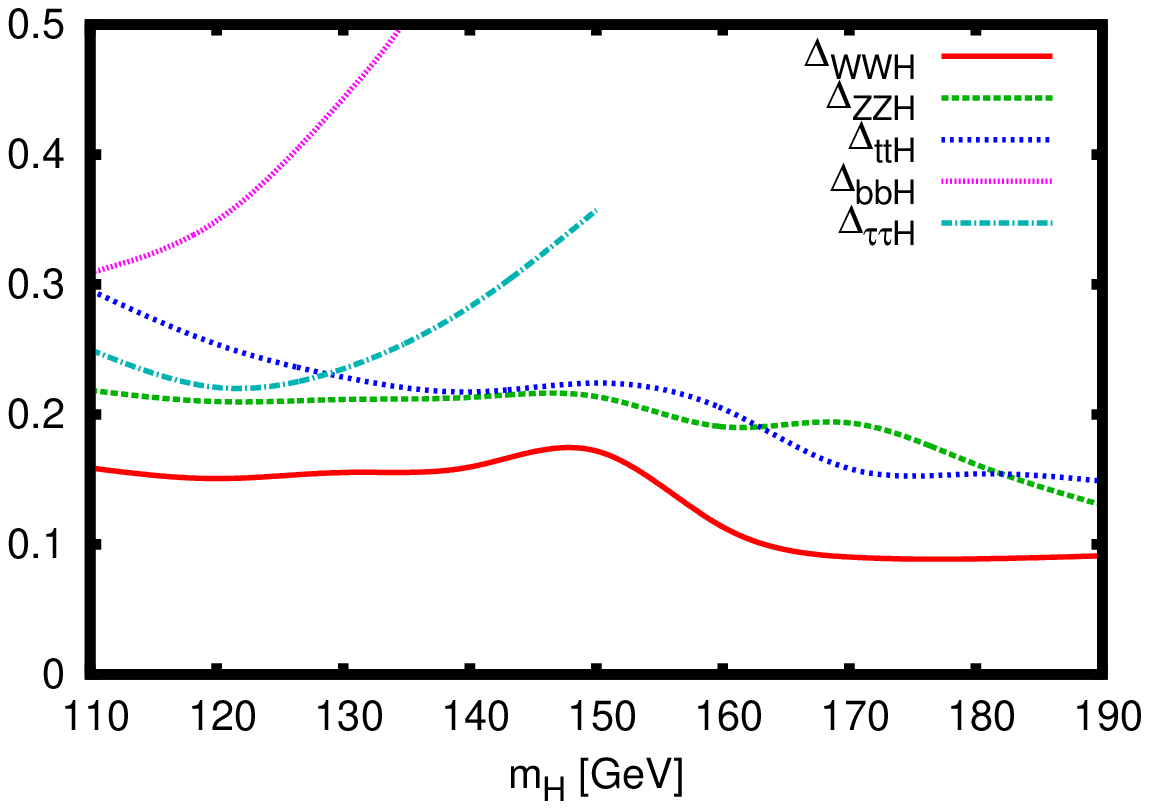,width=0.45\textwidth}
\psfig{figure=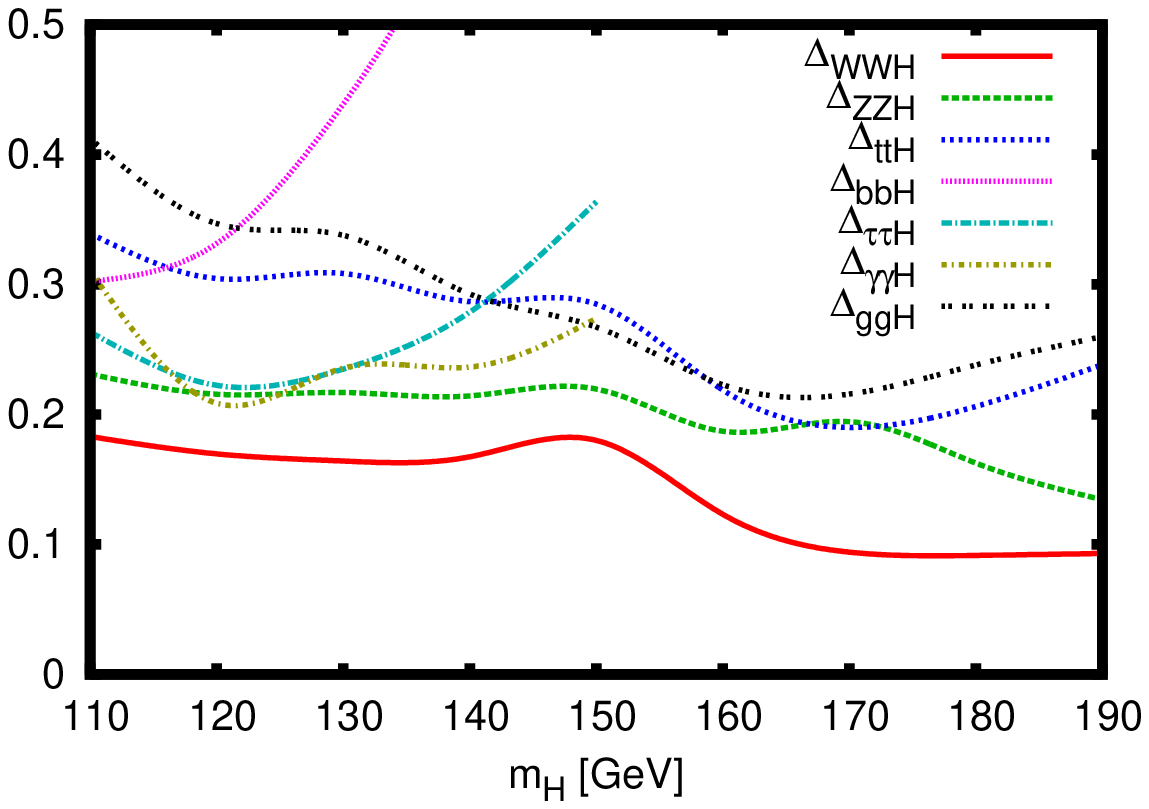,width=0.45\textwidth}
\vspace*{-3ex}
\caption{Error on the Higgs-boson couplings as a function of the Higgs
mass without (\textit{left}) and including (\textit{right}) additional
dimension-five operators for the 14 TeV LHC with 30 fb$^{-1}$ and SM
couplings.
\label{fig:sm_mh}}
\end{figure}

In Fig.~\ref{fig:sm_mh} we present the results of our analysis. We show
the 68\% CL errors on $\Delta_{jjH}$ for the different couplings. As
input we take the expected measurements at the given Higgs mass assuming
SM coupling strength. For other central values of the coupling, the
difference on the errors is small. On
the left-hand side we have neglected any contribution from additional
dimension-five operators to the effective couplings, while on the
right-hand side these are included as well. In both cases the $WWH$
coupling is the best determined one, with an error ranging from 10 to 20
\%. The region around 150 GeV is thereby the most difficult, because the
rates for processes with Higgs decays into bottom quarks are already
small and make these measurements difficult, as can be seen from the
error on $\Delta_{bbH}$. Nevertheless it is still
large enough to give a significant contribution to the total width,
thereby influencing the branching ratios to all other particles.
A large difference between the plots occurs for the top-quark Yukawa
coupling. Without the additional contributions, this is mainly
determined by its dominant role in gluon-fusion Higgs production.
Including it, a shift can always be compensated by a corresponding
contribution of the dimension-five operator. Only the channels with
top-quark associated Higgs production can determine the size of the
$ttH$ coupling.

\begin{figure}
\centering
\psfig{figure=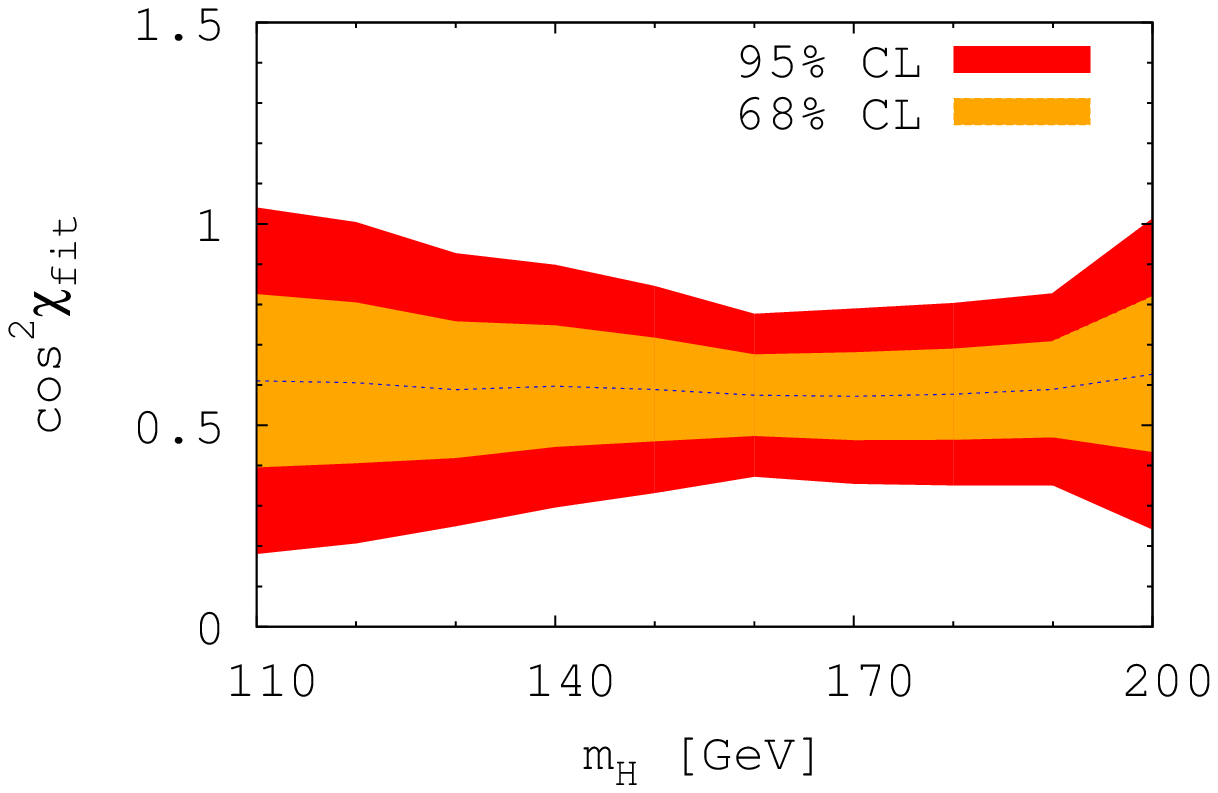,width=0.45\textwidth}
\psfig{figure=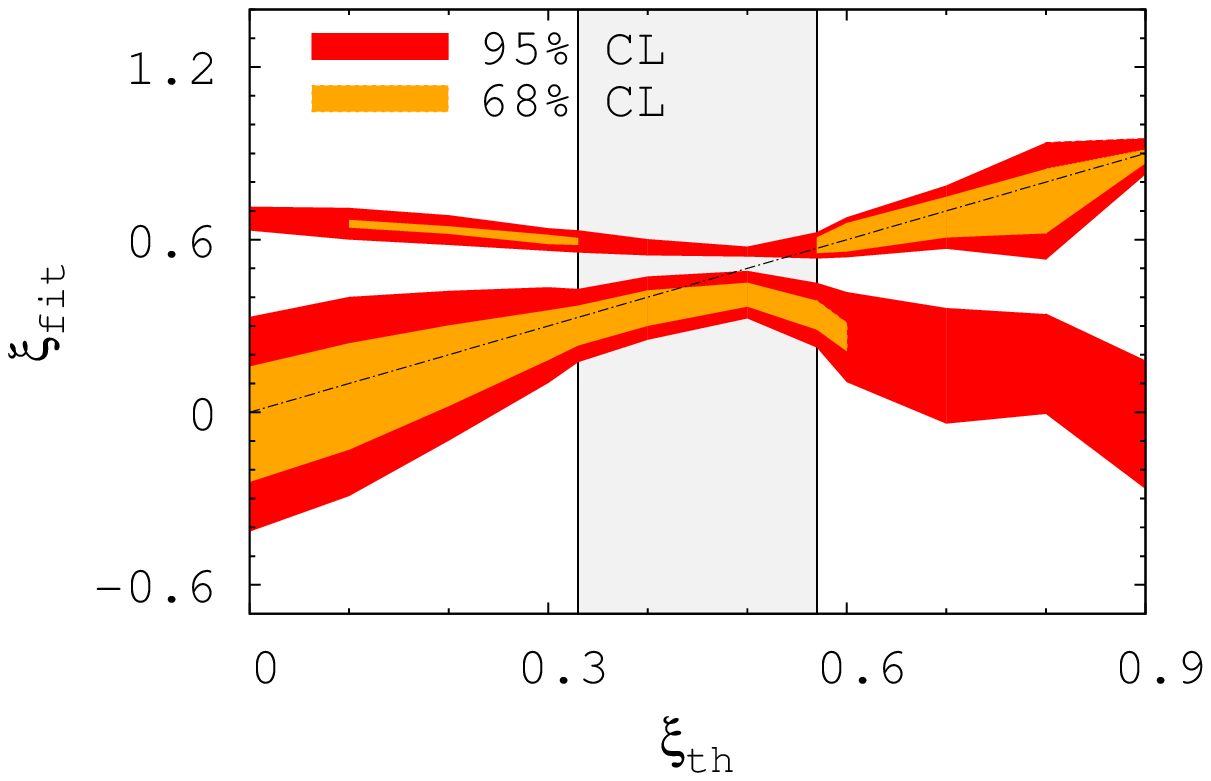,width=0.45\textwidth}
\vspace*{-3ex}
\caption{Obtainable precision of the fitted parameter at the LHC with 14
TeV and 30 fb$^{-1}$. \textit{Left:} Higgs portal model with input value
$\cos^2\chi_\text{th}=0.6$. \textit{Right:} SILH model MCHM 5 at $m_H = 120$
GeV.
Both figures from Ref.~\protect\cite{Bock:2010nz}.
\label{fig:portal}}
\end{figure}

Once we go to new-physics models, the size of the Higgs couplings, which
is predicted by the theory, can differ from the SM
value~\cite{Bock:2010nz}. Here, we shortly discuss two different models,
the Higgs
portal~\cite{Patt:2006fw} and a strongly-interacting light
Higgs (SILH)~\cite{Espinosa:2010vn}. In the first model, an additional
hidden sector is added, which is a singlet under the SM gauge groups. The
only possible connection
to the SM is via a quartic coupling $\mathcal{L} \propto
\Phi_s^\dagger \Phi_s \Phi_h^\dagger \Phi_h$ of the SM Higgs $\Phi_s$
and the hidden sector Higgs field $\Phi_h$. This induces a mixing
between the two Higgs particles, reducing the coupling between Higgs
and SM particles by a factor $\cos\chi$. Additionally, decays of the
Higgs into hidden sector particles appear if they are kinematically
allowed, and lead to invisible Higgs decays. This leads to simple
one-parameter fits such as the one shown in Fig.~\ref{fig:portal} on the
left. Here we show for a theory input of $\cos^2
\chi_{\mathrm{th}}=0.6$ and $\Gamma_\text{hid}=0$ the resulting fit for
$\cos^2\chi$ as a function of the Higgs mass, assuming 30 fb$^{-1}$
collected at 14 TeV. We see that in this scenario the SM
($\cos^2\chi=1$) can be excluded at the 95\% CL over the whole mass
range.

In the SILH model, the Higgs emerges as a pseudo-Goldstone boson of a
new strongly-interacting sector. The modifications of the couplings are
parametrised by $\xi = (\tfrac{v}{f})^2$, where $f$ is the Goldstone
scale. 
This model also predicts significant deviations for Higgs pair
production, which we do not consider in this analysis~\cite{Grober:2010yv}.
In the so-called MCHM4 scenario all Higgs couplings scale as
$\sqrt{1-\xi}$, so we can take the previous results by identifying
$\cos^2 \xi = 1-\xi$ and setting invisible decays to zero.
In the MCHM5 scenario, the vector-boson Higgs couplings are modified as
above, but the fermion couplings receive the factor
$\tfrac{1-2\xi}{\sqrt{1-\xi}}$. This leads to interesting coupling
structures, as the fermion couplings vanish for $\xi=0.5$ and rise
again for smaller values with opposite sign. On the right side of
Fig.~\ref{fig:portal} we show the results for the fitted value of $\xi$
over the theory input. We see that for each input value two
solutions emerge. For example, in the gluon fusion channel
with Higgs decays into photons, for each theory value of $\xi$
there are two possible fitted solutions that give the same rate. For
vector-boson-associated production with decays into bottom quarks using
subjet techniques~\cite{subjet}, another important channel, multiple
solutions only exist for values of $\xi \gtrsim 0.4$. Therefore, this
degeneracy is not a true one, but is induced by statistical fluctuations in
the measurements.  Each of the smeared toy experiments has a unique
solution, but whether it is the correct one can only be resolved by higher
integrated luminosities.

\section*{Acknowledgements}
We would like to thank the organisers of the ``23rd Rencontres de Blois
2011'' for the inspiring atmosphere during the workshop. This research
was supported by the Deutsche Forschungsgemeinschaft via the
SFB/TR-9 ``Computational Particle
Physics'' and the Initiative and Networking Fund of the Helmholtz
Association, contract HA-101 (``Physics at the Terascale'').

\end{document}